\documentclass[twocolumn,preprintnumbers,aps,showpacs,superscriptaddress,pra]{revtex4-1}

\usepackage{palatino}
\usepackage[latin1]{inputenc}
\usepackage{epsf}
\usepackage{amsmath,amssymb}
\usepackage{latexsym}
\usepackage{calc}
\usepackage{color}
\usepackage{shadow}
\usepackage{epsfig}
\usepackage{float}

\usepackage[pdftex, pdfstartview = {FitH}]{hyperref}

\newcommand{\ben}{\begin{equation}}
\newcommand{\een}{\end{equation}}
\newcommand{\bea}{\begin{eqnarray}}
\newcommand{\eea}{\end{eqnarray}}

\def\dulr{{\underline{\underline{\bf r}}}}

\begin{document}
\title{Machine Learning Exchange-Correlation Potential in Time Dependent Density Functional Theory}

\author{Yasumitsu Suzuki}
\email[]{yasumitsu.suzuki@rs.tus.ac.jp}
\affiliation{Department of Physics, Tokyo University of Science, 1-3 Kagurazaka, Shinjuku-ku, Tokyo 162-8601, Japan}  
\author{Ryo Nagai}
\affiliation{Department of Physics, The University of Tokyo, Hongo, Bunkyo-ku, Tokyo 113-0033, Japan} 
\affiliation{Institute for Solid State Physics, The University of Tokyo, Kashiwa, Chiba 277-8581, Japan} 
\author{Jun Haruyama}
\affiliation{Institute for Solid State Physics, The University of Tokyo, Kashiwa, Chiba 277-8581, Japan}

\date{\today}

\begin{abstract}
We propose a machine learning based approach to develop the exchange-correlation potential of time dependent density functional theory (TDDFT).
The neural network projection from
the time-varying electron densities to the corresponding 
 correlation potentials in the time-dependent Kohn-Sham equation
is trained using a few exact datasets for a model system of electron-hydrogen scattering. 
We demonstrate that this neural network potential 
can capture the complex structures in the time-dependent correlation potential 
during the scattering process 
and provide correct scattering dynamics, which are not obtained by the standard 
adiabatic functionals. We also show that it is possible to incorporate 
the nonadiabatic (or {\it memory}) effect in the potential with this 
 machine learning technique, which significantly improves the accuracy 
 of the dynamics. The method developed here offers a novel way to improve
 the exchange-correlation potential of TDDFT,
 which makes the theory a more powerful tool to study various excited state phenomena.
\end{abstract}

\maketitle

Time-dependent density functional theory (TDDFT)~\cite{tddft1,tddft2,tddft3} is 
a widely used first-principles approach 
to study the excited state properties of atoms, molecules and solids. 
TDDFT enables the first-principles simulation of correlated many-electron dynamics, which 
is in principle described by the time-dependent
Schr\"{o}dinger equation (TDSE) for the interacting system, by mapping it
to the dynamics of the noninteracting [also called Kohn-Sham (KS)] system
evolving in a single-particle potential.
There have been many successful applications of TDDFT simulation to 
the interpretation and prediction of various excited state 
phenomena, e.g., the linear response and spectra of molecules and solids~\cite{spectra1,spectra2,spectra3,spectra4,spectra5},
and real-time electron dynamics in systems exposed to external fields~\cite{dynamics1,dynamics2,dynamics3,dynamics4,dynamics5,dynamics6}
and in various non-equilibrium situations~\cite{neq0,neq1,neq2,neq3,neq4,neq5}.

TDDFT is a formally exact theory, i.e., it ensures that
the TDSE for the noninteracting (KS) system,
\ben
\begin{split}
i\frac{\partial}{\partial t}&\Phi_{\rm KS}(\dulr, t)
=( \sum_{i=1}^N [ -\frac{\nabla^2_i}{2}+ v_{\rm ext}({\bf r}_i,t) \\
&+ v_{\rm H}[n]({\bf r}_i,t) +v_{\rm XC}[n,\Psi_0,\Phi_0]({\bf r}_i, t) ] )\Phi_{\rm KS}(\dulr, t),
\end{split}
\label{eqn:tdkseq}
\een
can, in theory, yield any observables of an $N$-electron system exactly and solely from the time-dependent
electron density $n({\bf r}, t)=N \int d^{N-1}{\bf r} |\Phi_{\rm KS}(\dulr, t)|^2$. (Throughout this paper, 
atomic units are used unless stated otherwise, and 
$\dulr \equiv \{ {\bf r}_1,{\bf r}_2,\cdots,{\bf r}_{N} \}$.) 
Here, $v_{\rm ext}({\bf r}, t)$ and $v_{\rm H}({\bf r}, t)$ are 
the external potential applied to the system and 
the Hartree potential ($v_{\rm H}({\bf r}, t)=\int d{\bf r'}\frac{n({\bf r'}, t)}{|{\bf r}-{\bf r'}|} $), respectively,
and $v_{\rm XC}({\bf r}, t)$ is
the time-dependent (TD) exchange-correlation (XC) potential, which 
incorporates all many-body effects in the theory. 
The unique existence of the TDXC potential is proved by the Runge-Gross~\cite{tddft1} and van Leeuwen~\cite{Leeuwen} theorems; however,
its exact form is unknown. 
It is known that the
exact TDXC potential $v_{\rm XC}[n,\Psi_0,\Phi_0]({\bf r}, t)$ at time $t$, in principle, is functionally dependent 
on the history of the density $n({\bf r}, t'<t)$,
the initial interacting many-body state $\Psi_0$, 
and the choice of the initial KS state $\Phi_0$,
which indicates its exact form should be extremely complicated.

Therefore, almost all TDDFT applications to date 
use an adiabatic approximation,
which inputs the instantaneous density into 
one of the existing XC potential functionals in the {\it ground-state}  
density functional theory (DFT)~\cite{nonlocality1}, and completely
neglects both the history and initial-state dependence,
i.e., lacks the {\it memory effect}~\cite{tddft2,neepa1}.
It is true that the TDDFT calculation with these adiabatic functionals 
has achieved significant success in many studies~\cite{tddft4,tddft5,tddft6,tddft7,tddft8,tddft9,tddft10,tddft11,QSAC17}.
However, it has also been reported that
there are many situations where these approximate TDXC potentials
fail to even qualitatively reproduce the true dynamics~\cite{fail1,fail2,fail3,fail4}.
Recent studies on exactly-solvable model systems~\cite{xc1,xc2,xc3,xc4,RG12,EM12,yasu1,yasu2,helbig} have extensively explored the conditions where the adiabatic functional fails.
One important finding is that, when the local acceleration of 
electron densities occurs, the correlation part ($v_{\rm C}$) of the exact TDXC potential $v_{\rm XC}$ ($=v_{\rm X}+v_{\rm C}$), exhibits 
complex dynamical structures~\cite{xc1,yasu1} that arise from the memory effect and play significant roles to provide 
the correct dynamics.
The electron scattering process is a typical situation where these
complex structures in the TD correlation (TDC) potential appear, and 
it was revealed that the standard adiabatic functionals lack these structures~\cite{yasu1,yasu2}.

In this study, we propose   
a novel approach to improve the XC potential of TDDFT 
using a machine learning technique. 
Development of the 
XC functional in {\it DFT} by a machine learning based approach
has been actively conducted recently~\cite{ML1,ML2,ML3,nagai1,EXcondition,ML4,nagai2},
which demonstrates it is a promising direction to improve the DFT.
In particular,  
neural-network (NN) projection 
from the electron density to the ground-state XC potential
was developed in a recent study~\cite{nagai1},
and it was
demonstrated that the KS equation equipped with this NN functional
provides accurate ground-state density and total energy
for a one-dimensional two-electron model system, and has a remarkable transferability.
In TDDFT, the TDXC potential functional can also be regarded as a
projection $n\rightarrow v_{\rm XC}$, but here $n$ is the history of the density ($n({\bf r}, t'<t)$).
Thus, the projection should be more complicated than that in DFT, which means that 
there is more expectation on a machine learning based approach to find such a complicated projection.  
 
Here we construct the NN projection from the TD
density $n({\bf r}, t)$ to the TDC potential $v_{\rm C}({\bf r}, t)$ for
 a model system of electron-hydrogen (e-H) scattering~\cite{yasu1,yasu2}, 
as one example where the existing approximate functionals fail to 
reproduce the complex structures in $v_{\rm C}({\bf r}, t)$.
We demonstrate that 
this NN TDC potential $v_{\rm TDC}^{\rm NN}$ captures the 
complex structures that appeared in the exact potential very well,
and provide significantly improved time-resolved scattering dynamics 
compared to those obtained by the standard approximate functionals.

\begin{figure}[h]
 \centering
 \includegraphics*[width=1.0\columnwidth]{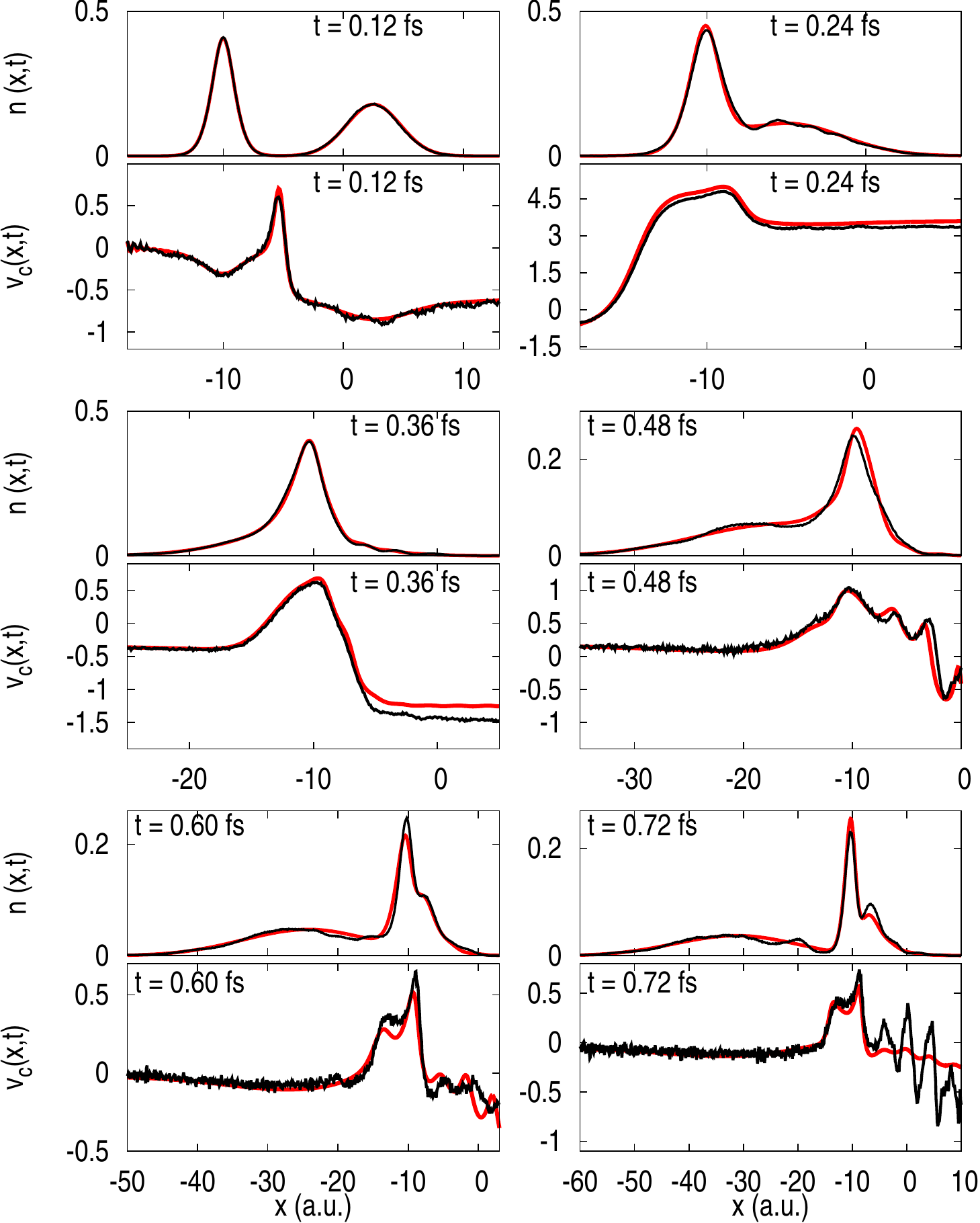}
 \caption{Snapshots of the trained NN TDC potential $v_{\rm TDC}^{\rm NN}$
for the initial KS state $\Phi_0^{(1)}$ in the e-H scattering model system
 (black line in the lower panel for each time slice)
and the corresponding TD electron densities (black line in the upper panel) obtained by
propagating the TDKS equation equipped with $v_{\rm TDC}^{\rm NN}$.
Exact electron density 
 (red line in the upper panel) and 
the exact TDC potential $v_{\rm TDC}^{\rm exact}$ (red line in the lower panel) are also plotted for each time slice.
}
 \label{fig:Fig1}
\end{figure}

The e-H scattering model system studied in this work is the same as that used in the previous studies~\cite{yasu1,yasu2}.
It is a one-dimensional two-electron system with the Hamiltonian:
$
{\hat H}(x_1,x_2)=\sum_{i=1,2}\left(-\frac{1}{2}\frac{\partial^2}{\partial x_i^2}+v_{\rm ext}(x_i)   \right) +  W_{ee}(x_1,x_2)
$,
where  $W_{ee}(x_1,x_2)=\frac{1}{\sqrt{(x_1-x_2)^2+1}}$ is the 
soft-Coulomb interaction~\cite{soft,soft1,soft2,soft3,soft4,soft5,soft6}
and the external potential
 $v_{\rm ext}(x)=-\frac{1}{\sqrt{(x+10)^2+1}}$ is
the soft-Coulomb model of a H atom located at $x=-10.0$ a.u.
The spatial part of the initial interacting wavefunction is 
$
\Psi_0(x_1,x_2) = \frac{1}{ \sqrt{2}}\left(\phi_{\rm H}(x_1)\phi_{\rm WP}(x_2) +\phi_{\rm WP}(x_1)\phi_{\rm H}(x_2) \right)
$, 
where a singlet state is chosen for the spin part.
 $\phi_{\rm H}(x)$ is the ground-state of
one electron alone in the external potential $v_{\rm ext}(x)$ and
$
\phi_{\rm WP}(x)=\left(2\alpha/\pi\right)^{\frac{1}{4}}e^{\left[-\alpha(x-x_0)^2+ip(x-x_0)\right]}
$
is an incident Gaussian wavepacket ($\alpha=0.1$), 
which represent an electron initially localized at $x_0=10.0$ a.u. approaching the H atom with a certain momentum $p$.
For this system,
the full TDSE 
$i\partial_t\Psi(x_1,x_2,t)={\hat H}(x_1,x_2)\Psi(x_1,x_2,t)$
can be numerically solved exactly, and
 the resulting TD density (for the case of incident momentum $p=-1.5$ a.u. for our first example),
 which were already reported in Ref.~\cite{yasu1}, are plotted as the red lines in the upper panel for different time slices 
in Fig.~\ref{fig:Fig1}. 
As reported previously ~\cite{yasu1}, for this case of $p=-1.5$ a.u., the scattering is inelastic,
and some part of the wavepacket is reflected back after the 
collision at around 0.24 fs.

For this two-electron dynamics, the exact TDXC potential
can be numerically obtained for any choice of the 
valid initial KS state that satisfies the van Leeuwen theorem~\cite{xc3,EM12}.
Here, we focus on one natural choice
for the initial KS state, i.e., the Slater determinant~\cite{yasu1,EM12}:
$
\Phi_0^{(1)}(x_1,x_2)
=\phi_0(x_1)\phi_0(x_2) 
\label{eqn: Phi_2}
$
with one doubly-occupied spatial orbital
$\phi_0(x)=\sqrt{\frac{n_0(x)}{2}}\exp\left[i\int^x \frac{j_0(x')}{n_0(x')}dx'\right]$,
where $n_0$ and $j_0$ are respectively the initial density and current density of the interacting system. 
For this initial KS state $\Phi_0^{(1)}$, 
the exact $v_{\rm X}(x,t)$ and $v_{\rm C}(x,t)$ can be numerically calculated~\cite{tddft2,xc1,gfpim1,gfpim2}
using the exact TD density $n(x,t)$ and current density $j(x,t)$ obtained from the solution of the TDSE.
The numerically obtained $v^{\rm exact}_{\rm C}(x,t)$ 
(shown as the red lines in the lower panels of Fig.~\ref{fig:Fig1}) 
exhibit complex peak- and valley-like structures
that are crucial for scattering~\cite{yasu1}.

In this study, we aim to make the NN
learn this exact TD {\it correlation} potential $v^{\rm exact}_{\rm C}(x,t)$ because
the exact functional form of $v^{\rm exact}_{\rm X}(x,t)$ ($=-\frac{1}{2}\int dx' W_{ee}(x',x) n(x', t)$)
is already known for the system under focus~\cite{tddft2,xc1}. 
The structure of the NN TDC potential $v_{\rm TDC}^{\rm NN}$ constructed here 
is expressed as:
\ben
\begin{split}
{\bf v}^{\rm NN}_{\rm TDC}=\cdots f[W^{(2)}f[W^{(1)}{\bf n}+{\bf b}^{(1)}]+{\bf b}^{(2)}]\cdots,
\end{split}\label{eqn: vNN}
\een
where ${\bf v}^{\rm NN}_{\rm TDC}$ and {\bf n} are the vectorized representations 
of $v_{\rm TDC}^{\rm NN}$ and $n$
 ($f$ is a non-linear activation function (ReLU function~\cite{relu} here),
and $W^{(l)}$ ($l=1,2,\cdots$) and ${\bf b}^{(l)}$ are the weight matrices and bias vectors, of which
the components are optimized to minimize the training error).
As with the study on the NN XC potential in DFT~\cite{nagai1}, the form of Eq.~(\ref{eqn: vNN})
is, in principle, sufficiently flexible to be a numerically exact TDXC potential.
The input vector ${\bf n}$ should ideally represent the entire history of the density ($n({\bf r}, t'<t)$); however,
in the first example, the instantaneous density $n({\bf r}, t)$ is used as ${\bf n}$.

The training procedure of the first example is as follows. 
First, the learning data set (${\bf n}^{(i)}(t_j)$, ${\bf v}^{(i)}_{\rm TDC}(t_j)$)
is generated from the numerical calculation of $n(x,t)$ and $v^{\rm exact}_{\rm C}(x,t)$. 
Here, $i$ is the index that corresponds to the different 
scattering dynamics calculation with a different initial incident momentum, $p$.
 In the first example, five different initial momenta; $p=-1.0$, $-1.2$, $-1.4$, $-1.6$, $-1.8$
 were employed to generate the training data set (thus $i=1,\cdots,5$).
For each calculation with different $p$, the TDSE was numerically propagated 
with the discrete time step $\Delta t=2.4\times10^{-3}$ fs
up to $t=0.72$ fs, which corresponds to 300 time steps, and
thus $j=1,\cdots,300$.
Therefore, the $5\times 300=1500$ data set of (${\bf n}^{(i)}(t_j)$, ${\bf v}^{(i)}_{\rm TDC}(t_j)$) was generated.
${\bf n}^{(i)}(t_j)$ and ${\bf v}^{(i)}_{\rm TDC}(t_j)$ are the vectors obtained by 
the real-space discretization of $n^{(i)}(x,t_j)$ and $v^{(i)}_{\rm TDC}(x,t_j)$, respectively, onto
common $N_r=1200$ uniform mesh points, i.e.,
${\bf n}^{(i)}(t_j)=\{ n^{(i)}(x_1,t_j),  n^{(i)}(x_2,t_j),  \cdots, n^{(i)}(x_{N_r},t_j)\}$
and ${\bf v}^{(i)}_{\rm TDC}(t_j)=\{ v^{(i)}_{\rm TDC}(x_1,t_j),  v^{(i)}_{\rm TDC}(x_2,t_j),  \cdots, v^{(i)}_{\rm TDC}(x_{N_r},t_j)\}$.

The parameters of the NN (Eq.~(\ref{eqn: vNN})) are then optimized with the generated data set 
using a similar method to that reported in Ref.~\cite{nagai1}: 
The fully connected NN with two hidden layers with 1200 nodes are used, and
the root mean squared error between ${\bf v}^{(i)}_{\rm TDC}(t_j)$
and those calculated from ${\bf n}^{(i)}(t_j)$ by the NN is minimized 
with the adaptive moment estimation method (Adam)~\cite{adam} algorithm
implemented in the {\it Chainer} package~\cite{chainer}.
The initial estimate of the weight parameters is randomly
generated and the optimization is stopped after 20000 epochs. 
Other details of the optimization are the same as those used in Ref.~\cite{nagai1}. This optimization
procedure is sufficient to provide a NN that gives excellent results, as detailed later.

Finally, the trained NN TDC potential $v_{\rm TDC}^{\rm NN}$
is implemented in the time-dependent Kohn-Sham (TDKS) equation (Eq.~(\ref{eqn:tdkseq}))
for the initial KS state $\Phi_0^{(1)}$ with an initial incident momentum
$p=-1.5$, which is out of the $p$ used for the training; that is, the test of the present NN is demonstrated by numerically integrating the TDKS equation for $\Phi_0^{(1)}$:
\ben
\begin{split}
i\frac{\partial}{\partial t}&\phi(x, t)
= [ -\frac{\nabla^2}{2}+ v_{\rm ext}(x,t) + v_{\rm H}[n](x,t)\\
&+v_{\rm X}[n](x, t) +v_{\rm TDC}^{\rm NN}(x, t)] \phi(x, t),
\end{split}
\label{eqn:tdkseq2}
\een
over time, where $n(x,t)=2|\phi(x,t)|^2$ is calculated and the TDC potential is obtained
from the NN ${\bf v}^{\rm NN}_{\rm TDC}$ on-the-fly at each time step, 
for the initial condition out of the training data set.

The resultant $n(x,t)$ and $v_{\rm TDC}^{\rm NN}(x,t)$
are plotted as black lines in the upper and lower panels of Fig.~\ref{fig:Fig1}.
It is evident that the black lines show similar structures to the exact ones (red lines);
in particular, $v_{\rm TDC}^{\rm NN}$ captures the 
complex structures of the exact TDC potential, and 
 the density dynamics reproduce the certain amount of reflection probability seen in the exact dynamics.
 Therefore, the machine learning based approach is 
 confirmed as effective for the numerical implementation of $v_{\rm TDC}$,
 at least for this first example.
 From Fig.~\ref{fig:Fig1}, $v_{\rm TDC}^{\rm NN}$ gradually exhibits spatially oscillating structures
 as times passes, which is due to the accumulation of small errors in the TD density, 
an intrinsic problem of the TD calculation that does not exist in the case of the NN potential for DFT.
Nevertheless, 
the TD densities obtained from $v_{\rm TDC}^{\rm NN}$ show rather smooth structures 
during the entire simulation time (up to $t=0.72$ fs). 
This is achieved by the effect of the kinetic energy operator in Eq.~(\ref{eqn:tdkseq2}) 
as a regulator of the artificial oscillation, as with that for the DFT case~\cite{nagai1}.

\begin{figure}[h]
 \centering
\includegraphics*[width=1.0\columnwidth]{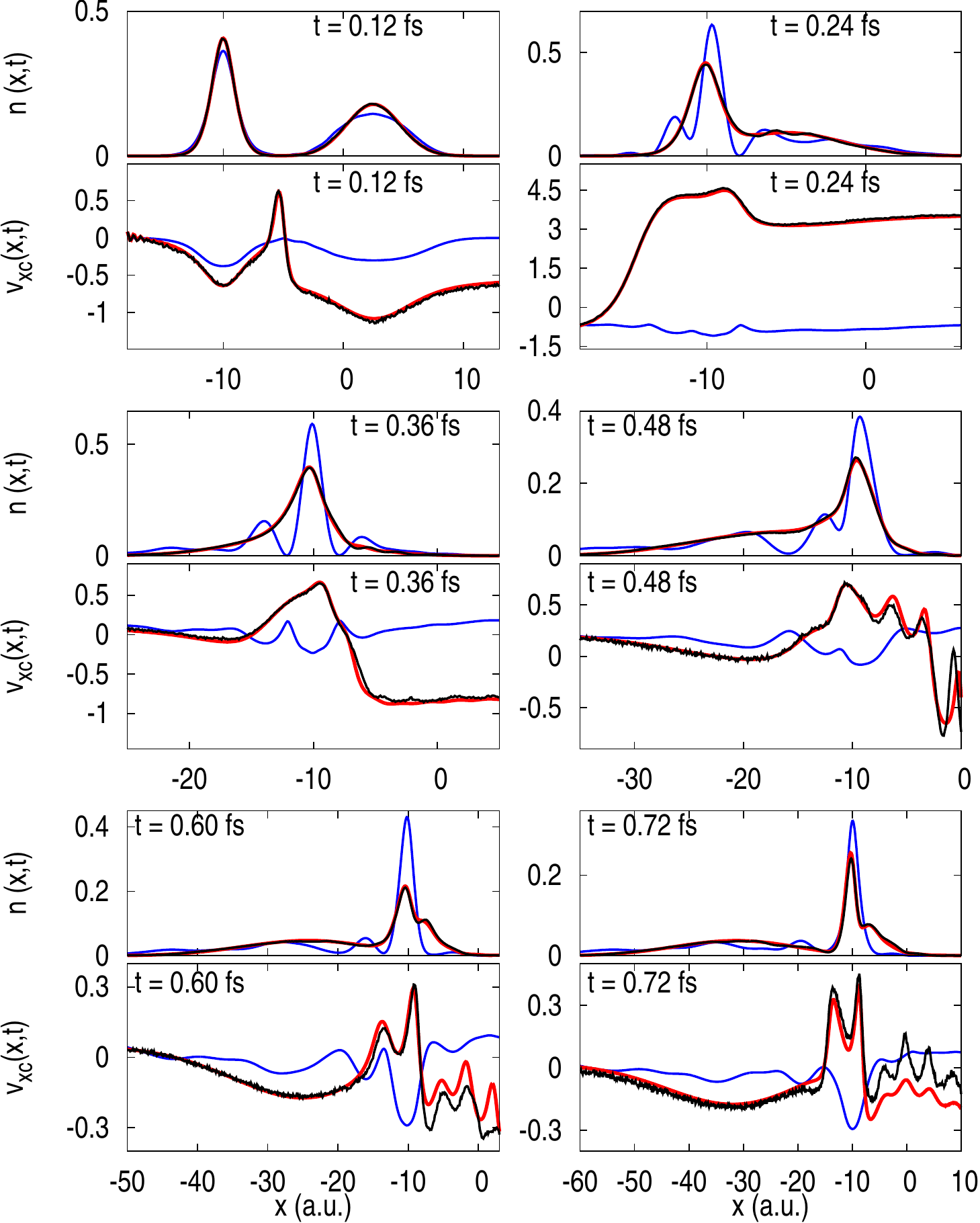}
 \caption{Snapshots of the 
 NN TDXC potential 
 with the memory effect
 $v_{\rm TDXC}^{\rm NN memory}=v_{\rm X}+v_{\rm TDC}^{\rm NN memory}$
for the same system shown in Fig.~\ref{fig:Fig1}
 (black line in the lower panel for each time slice)
and the corresponding TD electron densities  (black line in the upper panel).
Exact (red lines) and ALDA (blue lines) results are also plotted for each time slice.
}
  \label{fig:Fig2}
\end{figure}

Now we consider a strategy for improvement of  $v_{\rm TDC}^{\rm NN}$. 
Our first attempt to develop  $v_{\rm TDC}^{\rm NN}$ does not 
take account of the memory effect of $n({\bf r}, t'<t)$ explicitly, i.e., the training data set was the combination of the instantaneous density $n(x,t)$ and 
$v^{\rm exact}_{\rm TDC}(x,t)$.
Here we present how to incorporate the memory effect into $v_{\rm TDC}^{\rm NN}$.
We assume that the density distribution immediately before $t$ has the most effect on
$v_{\rm TDC}$ at $t$. 
Based on this hypothesis, we propose the following expression for the 
input vector {\bf n} for the NN (Eq.~(\ref{eqn: vNN})), so that it takes account of the memory effect:
\ben
{\bf n}^{(i)}_{\{t_0,t_j\}}=\{ {\bf n}^{(i)}(t_j), {\bf N}^{(i)}(t_j)\},
\label{eqn: memory}
\een
where ${\bf N}^{(i)}(t_j)$ is the vector representation of
\ben
N^{(i)}(x,t_j)=\int^{t_j}_{t_0}dt' w \left( |t'-t_j| \right)n(x,t'),
\label{eqn:tdwda}
\een
and $w(t)$ is the weight function, for which 
we employed Gaussian function $w(t)=\frac{A}{\sqrt{2\pi\sigma^2}}\exp\left(-\frac{t^2}{2\sigma^2} \right)$.
This additional input $N^{(i)}(x,t_j)$ (Eq.~(\ref{eqn:tdwda}))
in principle contains the history of $n$ from the initial time $t_0$, and 
$w(t)$ gives weight to the previous densities such that the more recent density has the larger effect.
This idea can be regarded as a {\it time version} of the average density approximation (ADA)~\cite{wda}
(The spatial version of ADA is a well-established technique to develop XC functional in DFT).
${\bf n}^{(i)}_{\{t_0,t_j\}}$ is related to ${\bf v}^{(i)}_{\rm TDC}(t_j)$ as one learning data set, i.e., 
the NN TDC potential trained with the memory effect, $v_{\rm TDC}^{\rm NN memory}$,
maps ${\bf n}_{\{t_0,t_j\}}$ to ${\bf v}_{\rm TDC}(t_j)$
at each time $t_j$.

\begin{figure}[h]
 \centering
 \includegraphics*[width=1.0\columnwidth]{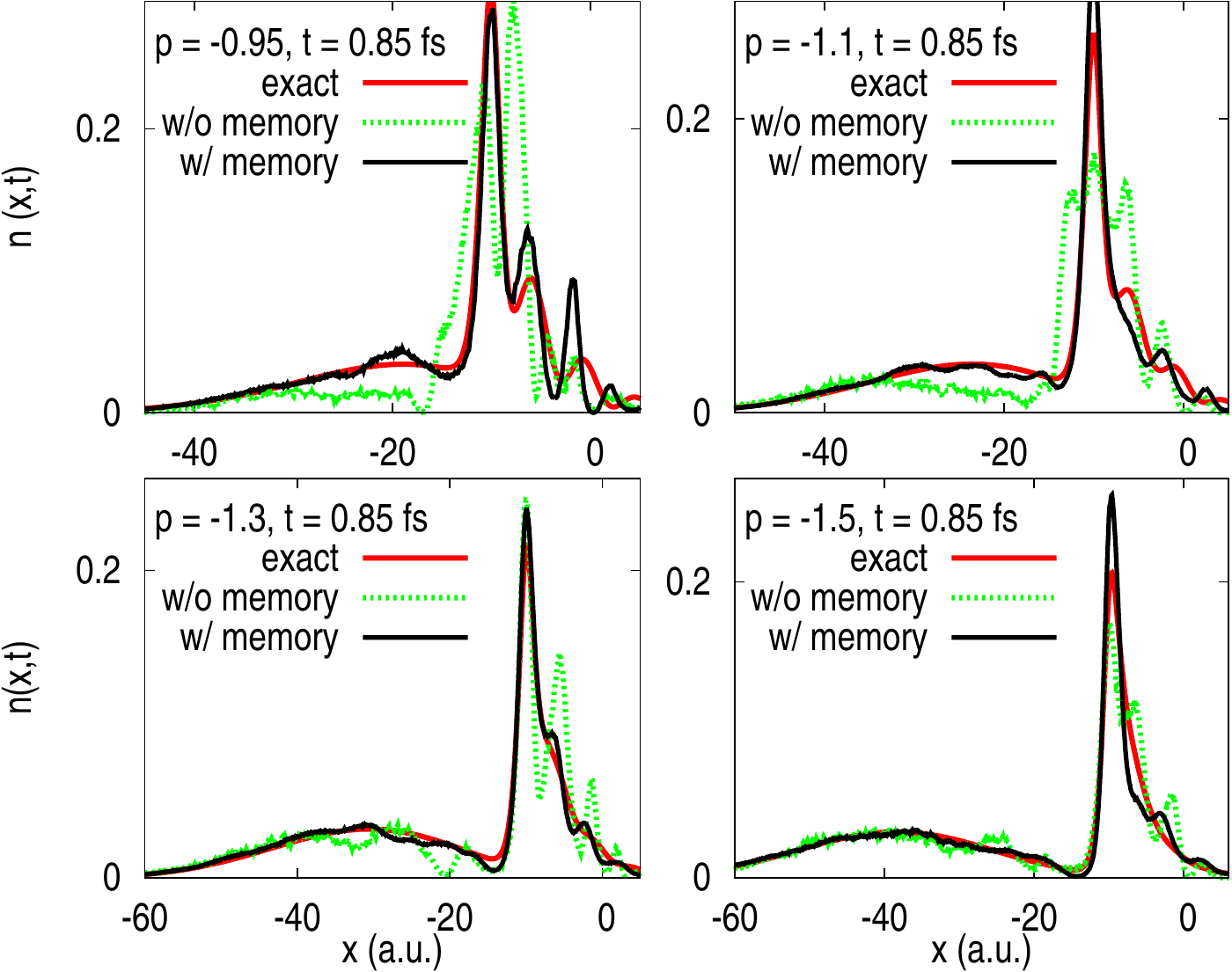}
 \caption{Snapshots of the electron density at 
  $t=0.85$ fs
   obtained from different calculations:
 the exact TDSE calculation (red solid line), the TDKS equation equipped with the NN TDC potential without the memory effect (green dotted line), and with the memory effect (black solid line). The four panels correspond to
 the different initial incident momentum 
 $p=-0.95$
 (upper left), $-1.1$ (upper right), $-1.3$ (lower left) ,
 and $-1.5$ (lower right). }
 \label{fig:Fig3}
\end{figure}

\begin{figure}[h]
 \centering
\includegraphics*[width=1.0\columnwidth]{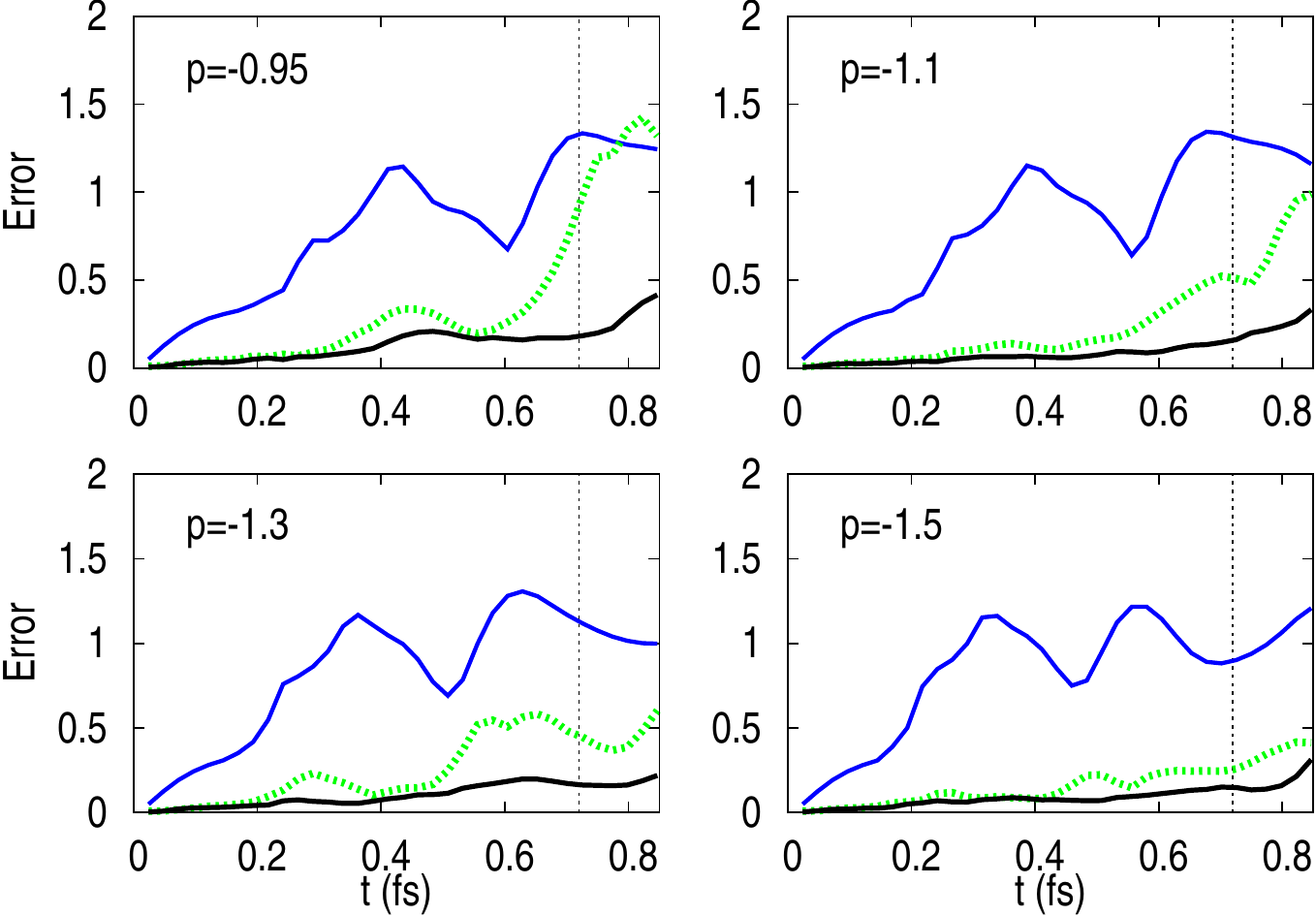}
 \caption{
 The error in the density for the
 NN TDC potential without the memory effect (green dotted line)
 and with the memory effect (black solid line), and ALDA (blue solid line)
 for the four different initial incident momenta; 
 $p=-0.95$, $-1.1$, $-1.3$,
 and $-1.5$ (indicated inside each panel). 
 Vertical dotted lines indicate
the simulation time for the training dataset ($0.72$ fs). 
}
  \label{fig:Fig4}
\end{figure}

We investigate the effectiveness of this method with $\sigma=2$ a.u. 
and $A=1$~\footnote[1]{We tested several different values for $\sigma$ and $A$, and
found they always give similar results discussed in the manuscript for the system under focus.}.
$v_{\rm TDC}^{\rm NN memory}$ is trained using a similar procedure to that without the memory effect
(The only difference is that the number of input nodes is now double, i.e., $1200\times2=2400$.
The number of hidden layer nodes is retained as $1200$).
Figure~\ref{fig:Fig2} shows 
snapshots of the NN TDXC potential with memory, i.e., $v_{\rm TDXC}^{\rm NN memory}=v_{\rm X}+v_{\rm TDC}^{\rm NN memory}$ (black solid line in the lower panels) and the TD density (black solid lines in the upper panels)
obtained through the solution of the TDKS equation with this NN potential.
A comparison of these results with the exact results (red lines) reveals remarkable agreement between them, and the results obtained from the NN with memory shows better agreement than those obtained without memory (Fig.~(\ref{fig:Fig1})); this 
is presented more clearly in Figs.~\ref{fig:Fig3} and~\ref{fig:Fig4} discussed later. 
We note that the exact TDXC potential (red line in the lower panels of Fig.~(\ref{fig:Fig2}))
and the exact TDC potential (red line in the lower panels of Fig.~(\ref{fig:Fig1}))
have almost the same structure, which indicates that 
the contribution of the TDX potential is small.
It was previously reported that 
the dynamics calculated with only the exact TDX potential functional
fail to reproduce the correct scattering~\cite{yasu1,yasu2}; therefore,
it is essential 
to capture the features of the exact TDC potential correctly.
In Fig.~\ref{fig:Fig2},
the results obtained using the adiabatic local density approximation (ALDA)~\cite{1dlda1,1dlda2,1dlda3} to 
both the exchange and correlation potentials
are also plotted as blue lines (same as those reported in Ref.~\cite{yasu1}).
The ALDA XC potential, and other standard XC functionals (reported in Ref.~\cite{yasu1,yasu2}),
lack the important memory effect, and their structures are significantly different from the exact structure, which leads to a failure
 to yield the correct scattering. 
The NN TDC potential presented here, both with and without the memory effect, provides
significantly better results than those from the standard functionals.

To show the impact of incorporating the memory effect, 
and check the out-of-training 
transferability of the NN,
we plot a comparison of the electron density at 
$t=0.85$ fs
obtained from the different calculations in Fig.~\ref{fig:Fig3};
 the exact TDSE calculation (red solid line), the TDKS equation equipped with the NN TDC potential without the memory effect (green dotted line) and with the memory effect (black solid line) 
 for the four different dynamics that start with different
 initial incident momenta; 
  $p=-0.95$,
   $-1.1$, $-1.3$, and $-1.5$ (indicated inside each panel).
None of these $p$ values are referenced in the training of the NN.
In particular, $p=-0.95$ is outside the range of training dataset.
Furthermore, the simulation time for these test dynamics (0.85 fs) 
is longer than that for the training dataset (0.72 fs).
Therefore, the out-of-training 
transferability of the NN, 
both for the parameter of the system and the simulation time,
can also be checked from Fig.~\ref{fig:Fig3}.
The results indicate that the NN potential with memory 
well reproduces the exact density at the time outside the training dataset for all $p$ cases.
Remarkably, it even reproduces the exact density for $p=-0.95$, which is outside the range of $p$ used for the 
training.
Thus, the out-of-training 
transferability of the NN potential with memory has been demonstrated. 
On the other hand, the density from the NN without memory 
has worse agreement with the exact density than that from the NN with memory, especially 
after $t=0.72$ fs and for $p=-0.95$
(this is also confirmed by Fig.~\ref{fig:Fig4} ; see below), which indicates 
a part of the memory effect is taken into account by the addition of $N^{(i)}(x,t_j)$
into the input to the NN.

\begin{figure}[h]
 \centering
\includegraphics*[width=1.0\columnwidth]{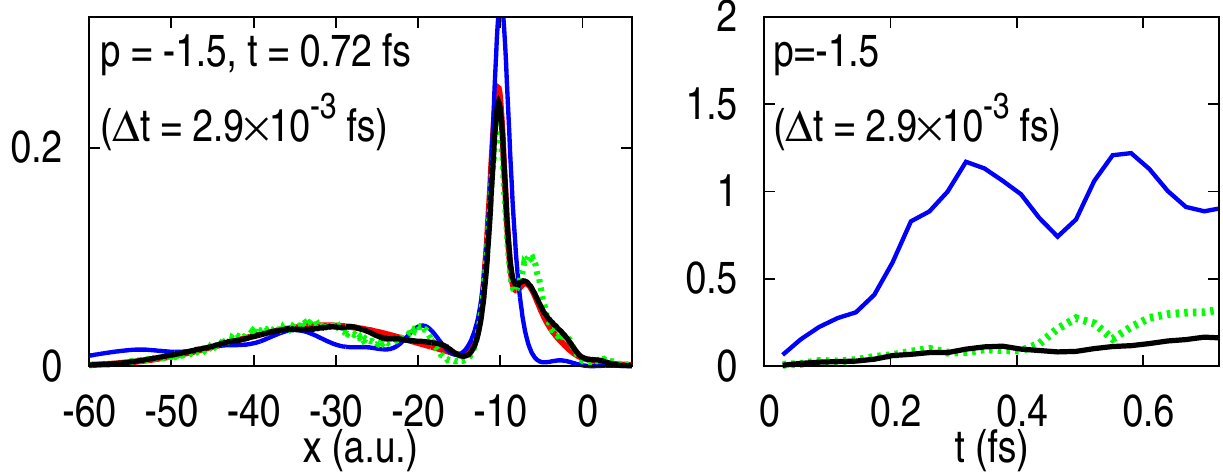}
 \caption{Snapshot of the electron density at $t=0.72$ fs for $p=-1.5$ obtained using $\Delta t=2.9\times10^{-3}$ fs
  from the exact (red solid), NN TDC potential without the memory effect (green dotted)
 and with the memory effect (black solid) and ALDA (blue solid) calculations
  (left panel). Right panel shows the corresponding errors in the density.
}
  \label{fig:Fig5}
\end{figure}

To clearly reveal the superiority of the NN with memory over that without memory, 
the time evolution of the deviation of the TDKS density from the exact one, which is defined as
$\int dx |n^{\rm exact}(x,t)-n^{\rm TDKS}(x,t)|$, is plotted
for all $p$ in Fig.~\ref{fig:Fig4} (green dotted line: the NN TDC
without memory, black solid line: the NN TDC with memory, and blue solid line: ALDA).
This figure confirms the present findings: 
the NN TDC with memory gives better results than the NN TDC without memory.
On the other hand, the ALDA gives poor results~\footnote[2]{It was previously reported~\cite{yasu1,yasu2} that ALDA is particularly 
problematic in the reproduction of the correct reflection.}.   
Therefore, the validity of the proposed strategy to incorporate the memory effect by
modification of the input density vector to Eq.~(\ref{eqn: memory}) is demonstrated.
We consider that the NN TDC potential with memory successfully captures
not only the space nonlocality~\cite{nonlocality1,nonlocality2}, which is important to match the exact TDX potential,
but also the time nonlocality (memory effect), at least for the model scattering problem investigated here.

We also point out the important advantage of the NN TDC potential using Eqs.~(\ref{eqn: memory}) and~(\ref{eqn:tdwda})
to take account of the memory effect; that is, this potential functional can be applied
to simulations with different $\Delta t$ from that used in the training dataset,
because the input $N^{(i)}(x,t_j)$ is calculated by integrating the previous densities over time.
Figure~\ref{fig:Fig5} shows the electron density at $t=0.72$ fs for $p=-1.5$ 
obtained from the calculations using $\Delta t=2.9\times10^{-3}$ fs
(left panel) and the corresponding errors in the density (right panel) 
  (Note that the training dataset is obtained using $\Delta t=2.4\times10^{-3}$ fs). 
  Again, the NN TDC potential with memory exhibits the excellent results, showing 
  its transferability to simulations with different size of time step.

We performed further test for the transferability of the NN with respect to the 
initial distance between the H atom and wavepacket and the parameter $\alpha$.
The results (shown in the supplementary information (Fig.~S1)) shows that the NN TDC potential 
has the transferability also with respect to these different parameters,
although the number of training dataset needs to be increased. 
This indicates that expansion of the region where the NN TDC potential can be applied requires
the expansion of the training data set.
This situation is similar to that of the NN for the DFT, where the possible characteristic states need to be included in the training data set to make the NN have a wider transferability~\cite{nagai1}.

Further improvement of the NN is expected by 
 enforcing the exact conditions of the TDXC potential,
 such as the zero force theorem~\cite{tddft2}, 
 in the NN~\footnote[3]{Violation of zero force theorem in the NN TDC potential here developed 
 is shown in supplemental materials (Fig. S2).}.  
 The exact conditions are important because 
 it ensure that the functional describes the correct physics,
 and recently it was employed to improve the NN functionals in DFT~\cite{EXcondition}.
  It could be possible to enforce the exact conditions in the NN for TDDFT as well,
  by using some specific functional form that satisfies the exact conditions
  regardless of the parameters optimized by the machine-leaning.
  This approach may also be used to suppress the oscillations in the NN potential.
The use of the recurrent neural network (RNN) ~\cite{RNN}
and long short term memory (LSTM)~\cite{LSTM} is also expected to show promise.

In summary, 
we have presented one example that indicates the novel approach based on the machine learning technique to develop the
XC potential of TDDFT is effective.
We have demonstrated that the NN TDC potential trained with a few 
numerically exact data sets reproduces the correct 1-dimensional two-electron scattering dynamics
that are not included in the training data, which demonstrates its transferability.
Furthermore, we have also shown that it is 
possible to incorporate the memory effect in this NN TDC potential, which significantly improves the result.

Our results indicate that once a few numerically exact (or sufficiently accurate) data of the
many-body dynamics of interest are available, then it is possible to train
 the NN TDC (or TDXC) potential so that it can be used to simulate at least similar dynamics to those
 used in the training.
  Applications of the NN potential to other systems where the memory effect is known to crucial, such as the system in an electric field~\cite{xc1}, is an 
  important feature direction 
  to investigate the need to further improve the functional form of NN to take account of the memory effect more effectively.
   The TDSE of an (three-dimensional) atom in a laser field
can be numerically solved by means of the time-dependent variational principle method~\cite{tdvp},
and the resulting data can be used to train the NN TDXC potential, which can then be applied to the TDDFT calculation
of actual molecules.
It will also be possible to apply the machine learning approach to develop the XC kernel
in linear-response TDDFT~\cite{spectra1,spectra2,spectra3,spectra4} 
using the many-body perturbation calculation results as training data.
With these studies, TDDFT will become a more powerful tool for the study of various excited state phenomena.

\begin{acknowledgments}
YS is supported by JSPS KAKENHI Grant No. JP19K03675.
Part of the computations were performed on
the supercomputers at the Institute for Solid State Physics,
The University of Tokyo.
\end{acknowledgments}

\bibliography{./reference}

\end{document}